\documentclass[article,aps,reprint,prb,twocolumn,amsmath,amssymb,superscriptaddress]{revtex4-1}%

\usepackage{graphicx}
\usepackage{lipsum}
\usepackage{relsize}
\usepackage{tikz}
\usepackage{dcolumn}
\usepackage{bm}
\usepackage{accents}
\usepackage{mwe}
\usepackage{natbib}
\usepackage{hyperref}
\usepackage{color}
\usepackage{amsmath}
\usepackage{epstopdf}
\usepackage{fancybox}
\usepackage{amsfonts}
\usepackage{amssymb}%
\usepackage[toc,page]{appendix}
\renewcommand{\cite}[1]{{[}\onlinecite{#1}{]}}

\newcommand{\be}{\begin{equation}}
\newcommand{\e}{\end{equation}}
\newcommand{\beml}{\begin{subequations}}
\newcommand{\eml}{\end{subequations}}
\newcommand{\beq}{\begin{eqnarray}}
\newcommand{\eq}{\end{eqnarray}}
\newcommand{\ba}{\begin{array}}
\newcommand{\ea}{\end{array}}
\newcommand{\bpm}{\begin{pmatrix}}
\newcommand{\epm}{\end{pmatrix}}
\newcommand{\bc}{\begin{cases}}
\newcommand{\ec}{\end{cases}}

\begin{document}

\title{Energy-efficient antiferromagnetic skyrmion creation and its dynamics in structure-dependent magneto-elastic coupling }
\author{Rohollah Khoshlahni}
\email{rkhoshlahn91@gmail.com}
\affiliation{Department of physics, Shahid Beheshti university, Tehran 19839, Iran}
\thanks{Corresponding author: rkhoshlahn91@gmail.com,m-mohseni@sbu.ac.ir}
\author{Yousef Azizi}
\affiliation{Independent Researcher, Zanjan 1111, Iran}
\author{Ali Aftabi}
\affiliation{Department of Physics, Faculty of Science, University of Kurdistan, Sanandaj 66177-15175, Iran}
\author{Serban Lepadatu}
\affiliation{Jeremiah Horrocks Institute for Mathematics, Physics and Astronomy, University of Central Lancashire, Preston PR1 2HE, United Kingdom}
\author{Majid Mohseni}
\email{m-mohseni@sbu.ac.ir}
\affiliation{Department of physics, Shahid Beheshti university, Tehran 19839, Iran}
\thanks{Corresponding author: m-mohseni@sbu.ac.ir}

\begin{abstract}
 Existing skyrmion nucleation methods lead to increased Joule heating, limiting the applicability to metallic Antiferromagnetic (AFM) systems. In this study, we propose a novel, energy-efficient mechanical method for nucleating AFM skyrmions using Surface Acoustic Waves (SAWs). SAWs, which propagate along material surfaces with minimal attenuation, generate dynamic strain that induces a spatially varying torque on magnetic spins, offering a viable alternative to traditional current-based methods. Using the Thiele approach we investigate skyrmion dynamics in special AFMs, NiO-like or CoO-like, incorporating a magnetoelastic term that accounts for the unique spin arrangements, where spins align oppositely within parallel planes. Our findings reveal that the deformation induced by SAWs, influenced by the magnetostriction of materials like NiO and CoO, can modify the spin configuration, consequently, alters the skyrmion dynamics. This study not only demonstrates the potential of SAWs for efficient skyrmion nucleation in AFMs but also introduces new theoretical insights into specific magnetoelastic-induced skyrmion dynamics in AFM systems.
\end{abstract}

\date{\today}
\maketitle
\section{Introduction}

Energy-efficient magnetic soliton creation and manipulation are two appealing issues in these days' spintronic science. Our methodology relies on surface acoustic waves (SAWs), which enable the creation and mechanical manipulation of magnetic textures through piezoelectric and magnetoelastic effects. The utilization of SAWs for the control of magnetic textures offers advantages such as low energy consumption and extended transmission distances \cite{chen2021surface}, with several studies documenting these efforts. The theoretical proposal for domain wall motion induced by SAWs was later validated through experimental work \cite{dean2015sound}. Recently, researchers employed a similar SAW device configuration to experimentally demonstrate the generation of skyrmions, being nanoscale swirling magnetic textures characterized by distinct topological properties \cite{moriya1960anisotropic,dzyaloshinsky1958thermodynamic}, by utilizing a spatiotemporally varying inhomogeneous effective torque \cite{yokouchi2020creation}.

By the way, AFM spintronics is a promising area of research due to several advantages, such as stability against external magnetic fields, ultrafast dynamics, the absence of stray fields, and low-power data transmission, making them suitable for spintronic devices\cite{neel1952antiferromagnetism, neel1936proprietes, jungwirth2018multiple, baltz2018antiferromagnetic, jungwirth2016antiferromagnetic,gomonay2014spintronics,marti2015prospect}.
 Sequentially, recent theoretical and experimental advancements in manipulating AFM spin textures, particularly domain walls (DWs), have renewed interest in their potential applications\cite{logan2012antiferromagnetic,wadley2018current,tveten2013staggered}.\\
 Until recently, skyrmions were exclusively observed in FM and long-wavelength spin spiral systems \cite{flovik2017generation, liu2015skyrmion, sampaio2013nucleation}. However, both theoretical \cite{zhang2016antiferromagnetic, keesman2016skyrmions, zhao2018single, bessarab2019stability, khoshlahni2019ultrafast} works and experimental observations \cite{legrand2020room, jani2021antiferromagnetic} suggest the possibility of stabilizing these topological objects, both as lattices and isolated entities, within AFM materials.\\
Skyrmion condensation and movement are worthwhile cases in data processing and transmission. Therefor, skyrmions can be created and displaced by various external perturbations, including electrical current \cite{iwasaki2013current, sampaio2013nucleation}, spin current \cite{montoya2018spin, kim2017current} ,temperature gradient \cite{khoshlahni2019ultrafast,gong2022dynamics, kong2013dynamics}, magnetic field \cite{komineas2015skyrmion}, anisotropy gradients \cite{shen2018dynamics} and surface wave \cite{nepal2018magnetic, yang2023magnetic, khoshlahni2023skyrmion}.\\
The proposed methods for creating and controlling AFM skyrmions face several limitations.
Many of these methods are currently applicable only to metallic AFM systems, limiting their use with other types of materials. Moreover, the proposed methods rely on the use of heterostructures, which can increase the complexity and difficulty of device fabrication. Furthermore, they require
relatively large current densities, which probably causes more Joule heating. These limitations underscore the need for further research and development to overcome these challenges and fully realize the potential of AFM skyrmions in future spintronic applications.\\
Here, we propose a low energy mechanical method to nucleate a  skyrmion in AFM systems. Magnets exhibit magnetoelastic coupling, a phenomenon where spin interactions are influenced by strain \cite{chikazumi1997physics}. Consequently, traveling strain waves, such as those generated by SAWs, induce a dynamic and spatially varying effective torque on the magnetic spins \cite{davis2010magnetization,  weiler2012spin, foerster2017direct, weiler2011elastically}. 
Importantly, SAWs propagate along the surface of a material with minimal attenuation, typically decaying over millimeter distances. Furthermore, they can be efficiently generated using electric fields through the inverse piezoelectric effect, avoiding the energy dissipation associated with Joule heating \cite{yokouchi2020creation}. 
Then, we will demonstrate skyrmion dynamics induced by SAW through Thiele approach in a special type of AFMs. These AFMs introduce a new crucial term that accounts for the specific arrangement of magnetic spins within the material. This term is particularly important for AfMs, where magnetic moments align in opposite directions within parallel planes. These planes are defined by a perpendicular axis \cite{simensen2018magnetostriction}. The study predicts that the material's deformation (magnetostriction) will change depending on the orientation of this perpendicular axis. These findings have significant implications for materials like NiO and CoO, where a novel type of strain directly influenced by the spin configuration has been observed \cite{simensen2019magnon}. \\
We classify the rest of this paper as follows. In Sec. \hyperref[II]{II} we declare our finding for the energy-efficient mechanical generation of isolated skyrmion in a typical AFM system. In Sec \hyperref[III]{III} the underlying spin interactions contained in free energy will be displayed. we will also study the micromagnetic free energy for CoO-like AFMs and introduce the additional magnetoelastic term. Then the main result will be declared, concisely. Theoretical Skyrmion dynamics and its numerical simulation are presented in Sec. \hyperref[IV]{IV}. In Sec. \hyperref[V]{V}, we present the results and outcomes of the study, and discuss promising directions for future work. Finally, the details of equation of motion calculations will be derived in Sec.\hyperref[VI]{VI} as appendix.
\section{Energy efficient mechanical nucleation of isolated skyrmion}\label{II}
\begin{figure}[t]
	\hspace{-0.9 cm}
	\begin{minipage}[b]{0.48\linewidth}
		\centering
		\includegraphics[width=1\linewidth]{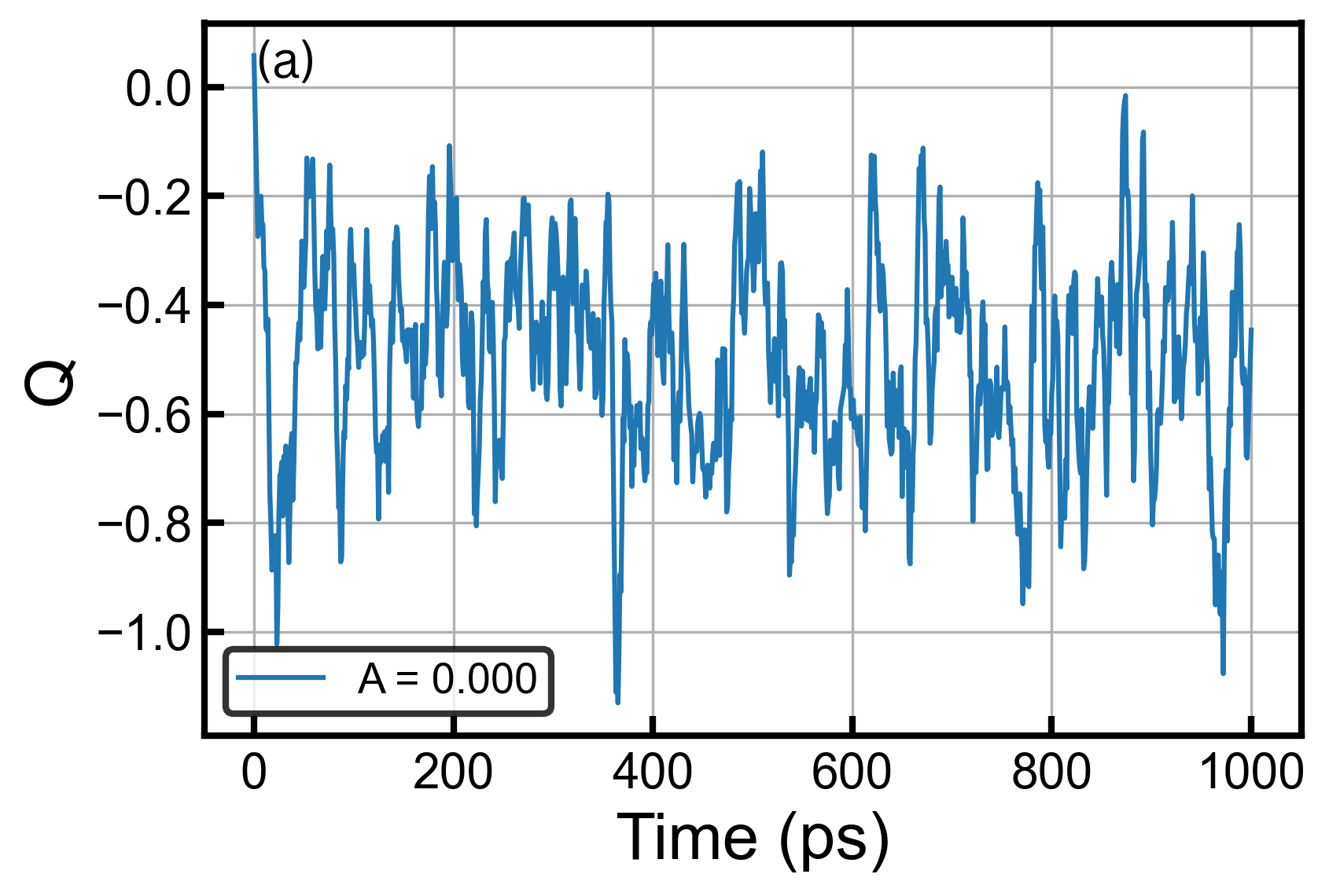}
 	\end{minipage}
	\hspace{0.5cm}
	\begin{minipage}[b]{0.48\linewidth}
		\centering
		\includegraphics[width=1\linewidth]{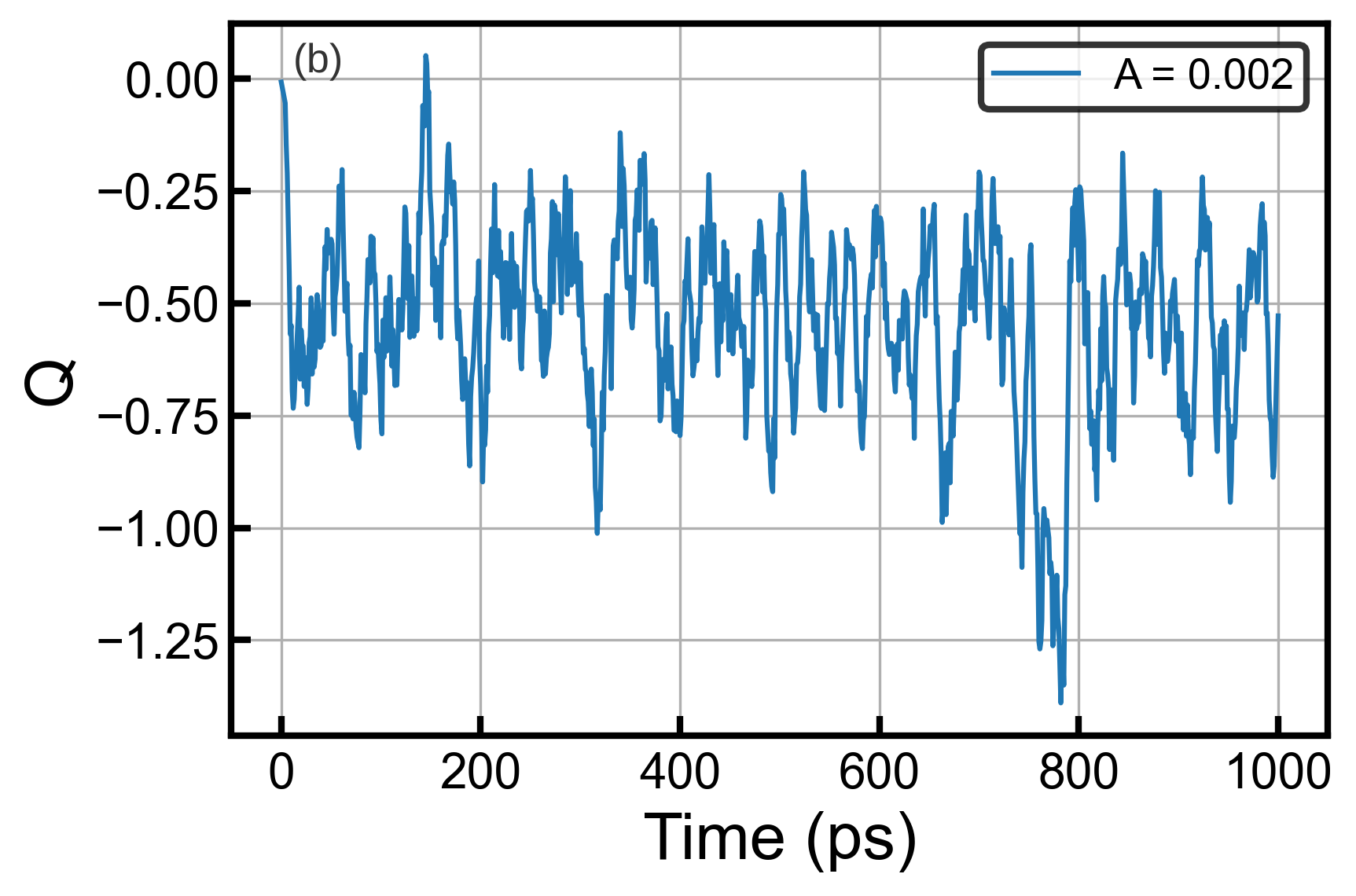}
	\end{minipage}
	\hfill
	\begin{minipage}[b]{0.48\linewidth}		
		\centering
		\hspace{-1.5 cm}
		\includegraphics[width=1\linewidth]{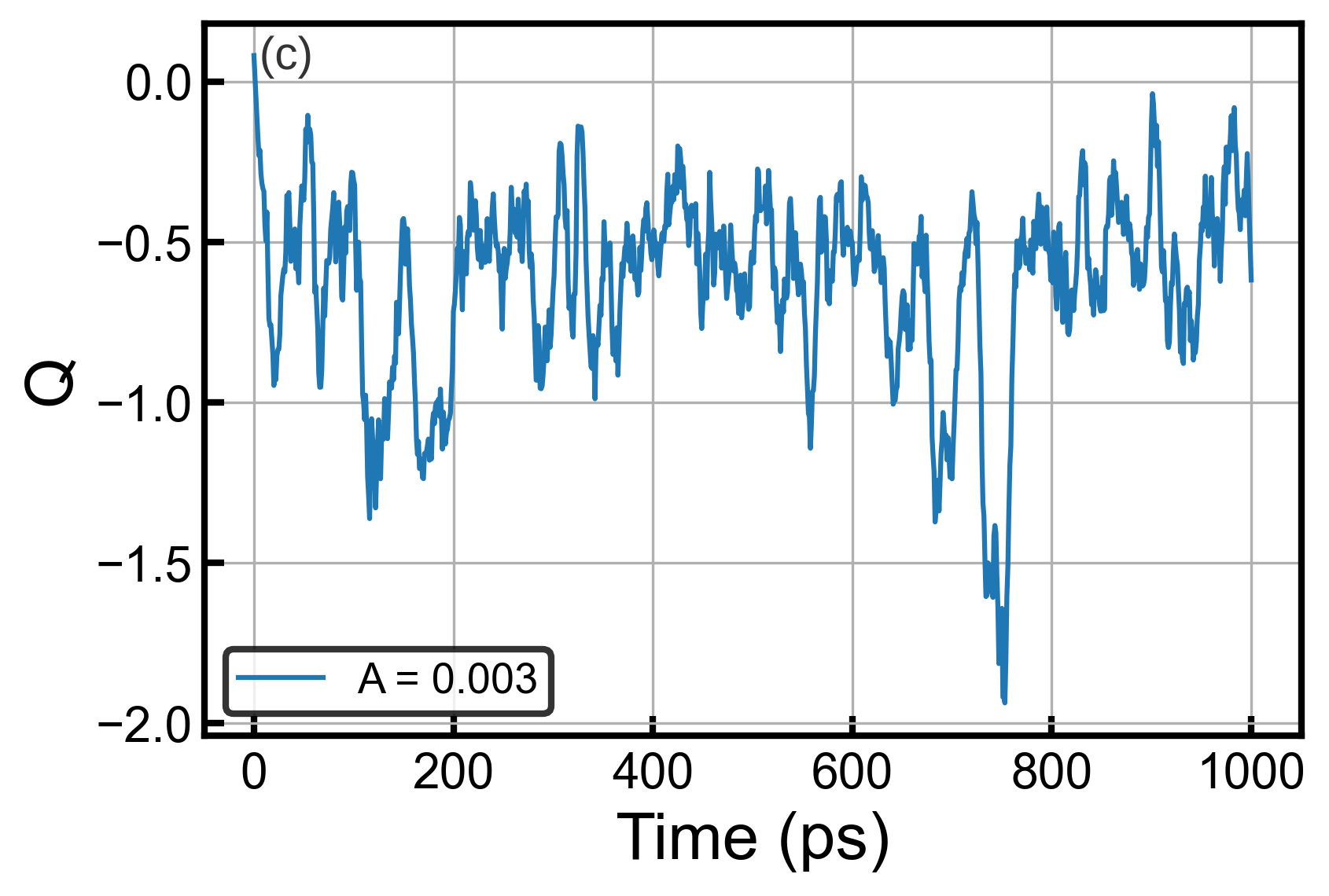}
	\end{minipage}
	\hspace{-0.3cm}
	\begin{minipage}[b]{0.48\linewidth}
		\centering
		\includegraphics[width=1\linewidth]{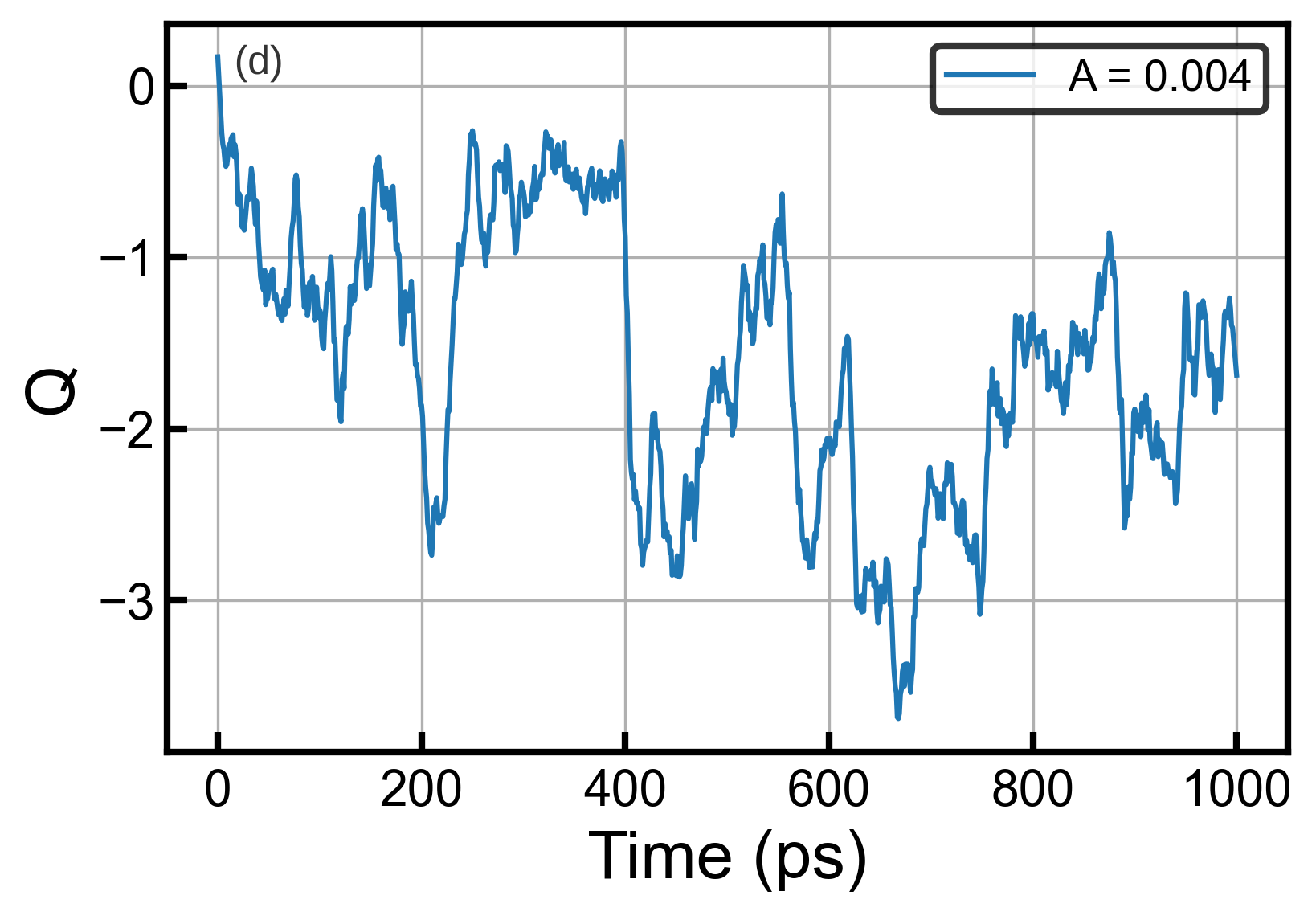}
	\end{minipage}
	\caption{ With A = 0 (or small) the ground state is uniform, i.e. no skyrmions.
	As the SAW amplitude A is increased skyrmions start to form - the larger the amplitude the more skyrmions form.
	 We have plotted the topological charge Q as a function of time for different amplitudes A, see in a-d.  		 For A = 0 and 0.002 there are no skyrmions formed (Q is noisy due to stochasticity and fluctuates around a background value of 0.5) (a,b). In (c), as A increases one could see spikes in the Q value, which is due to skyrmion nucleation events. For example at A = 0.003 single skyrmions tend to form, although more are possible.}
 \label{topo_number}
\end{figure}

\begin{figure}[!h]
\includegraphics[width=9.0cm,height=4cm,]{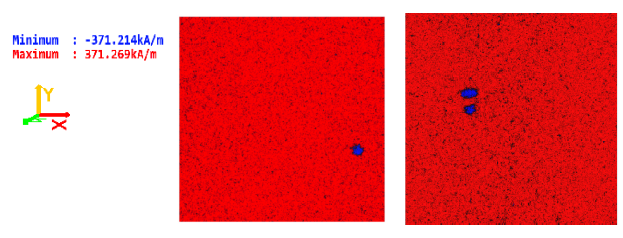}
	\caption{ At A = 0.003 single skyrmions tend to form (a), although more are possible. First at ~ 700 ps which shows an isolated skyrmion, and the other (b) at ~740 ps which shows two skyrmions (for A = 0.003) .}
	\label{sky_creation}
\end{figure}

Skyrmions exist either as a stable "skyrmion crystal" or as individual skyrmions in a metastable state. Isolated skyrmions are prominent in data storage and processing. Therefore, controlling individual skyrmions is essential for practical applications. In this section, we propose a mechanical method to create single skyrmions in confined geometries. Creating a single skyrmion in this less stable state involves transitioning the system from its ground state (a uniform magnetic alignment) to a new stable state containing a single skyrmion.\\
 We demonstrate, in FIG.\ref{topo_number} that when the SAW amplitude (A) is low or zero, the system remains in a uniform state without skyrmions. As the SAW amplitude (A) increases, skyrmions begin to form. Higher amplitudes generally lead to a greater number of skyrmions. Topological Charge (Q) in
small A fluctuates around 0.5, indicating the absence of skyrmions. On the other side, higher A spikes in the Q value appear, signifying the nucleation of skyrmions.
Single skyrmions can only form and disappear when there are random fluctuations in the system. These fluctuations come from vibrations within the material at non-zero temperatures, which always exist in real experiments.
Skyrmions are generated in one region where the surface acoustic wave (SAW) is present and subsequently annihilated in another region. These skyrmions are inherently unstable in the absence of the SAW. Finite temperature effects are incorporated through the inclusion of a standard stochastic term in the Landau-Lifshitz-Gilbert (sLLG) equation.\\
As shown in FIG.\ref*{sky_creation}, first at ~ 700 ps which shows two deformed skyrmions, and the other at ~740 ps which shows single skyrmions (for A = 0.003). The underlying mechanism for skyrmion creation is attributed to a modification of the effective magnetic anisotropy due to magnetoelastic coupling. This alteration effectively shifts the system into a parameter regime where skyrmion configurations become energetically favorable.
\\
\section{model and main results}\label{III}
\subsection{model}
The discrete magnetic free energy for a two-dimensional square magnetic system may be considered as:
\begin{align}
\mathfrak{F}[\mathbf{S}_i,\mathbf{S}_j]=&\sum_{<i,j>}J_{i,j}\mathbf{S}_i\cdot\mathbf{S}_j-\sum_{<i,j>}\mathbf{D}_{i,j}\cdot\mathbf{S}_i\times\mathbf{S}_j\nonumber\\
-&K\sum_{i}(S_z)^2-\sum_{i}\mathbf{H}\cdot\mathbf{S}_i+\mathcal{F}_{mec},
\end{align}
where $\mathbf{S}_{i,j}$ is the spin magnetic moment at sites $i,j$. The following terms are the nearest neighbor Heisenberg interaction with $J>0$, DMI with $D_{i,j}$ vector, uniaxial anisotropy with $K<0$, Zeeman and ME coupling, respectively. The exchange interaction with $J>0$ forces adjacent spins to be antiparallel with each other, the DMI encourages them to become perpendicular, magnetic anisotropy prefers the normal direction on a crystal plane, Zeeman effect and discrete magnetostriction energy, respectively. The long-range dipolar interaction can be neglected because in ultrathin magnetic system with no net magnetization the interaction may be ignored. 
Topological magnetic textures, such as DW, skyrmion,bimeron and hopfion \cite{zhang2016antiferromagnetic, woo2016observation,li2020bimeron,sapozhnikov2022creating,kent2021creation,wang2019current} are created through the competition of these energies both in ground states and metastable states. To illustrate the collective interactions inside the magnetic systems, it is reasonable to exploit the continuum model of spin system in which the lattice spacing tends to zero. Therefor the spins become homogeneous and may be substituted by magnetization, i.e $\mathbf{m}_{(\mathrm{A,B})}(\mathbf{r},t)$ for the sublattices $\rm{A,B}$.

Above the N\'{e}el temperature the material does not have antiferromagnetic order, on the other side,
bellow the N\'{e}el temperature the  staggered magnetization is $\mathbf{n}(\mathbf{r},t)=(\mathbf{m}_{\rm{A}}-\mathbf{m}_{\rm{B}})/|(\mathbf{m}_{\rm{A}}-\mathbf{m}_{\rm{B}})|$ in which $(\mathbf{m}_{\rm{A}}-\mathbf{m}_{\rm{B}})/2$ is the difference of sublattice's order parameter and $\mathbf{m}(\mathbf{r},t)=(\mathbf{m}_{\mathrm{A}}+\mathbf{m}_{\mathrm{B}})/2$  is the total magnetization for an AFM system where $|\mathbf{m}_{\rm{A}}-\mathbf{m}_{\rm{B}}|\approx 1 $ and $\mathbf{m}\ll\mathbf{l}$ so that these terms bring the
$\mathbf{n}\cdot\mathbf{m}=0$ and $\mathbf{n}^2+\mathbf{m}^2=1$ constraints. While in the equilibrium state $\mathbf{m}_{\mathrm{A}}$ and $\mathbf{m}_{\mathrm{B}}$ are antiparallel, $|\mathbf{m}_{\mathrm{A}}(\mathbf{r},t)|=|\mathbf{m}_{\mathrm{B}}(\mathbf{r},t)|=1$.

One can utilize these concepts and simplify the study of AFM system and go beyond the two-sublattice issue then receives at  staggered magnetization expression for total free energy.
Now we introduce the AFM free energy ingredients containing a novel term in ME interaction,
\begin{align}\label{eq:2 AFM free energy}
\mathfrak{F}[\mathbf{n}]=\int dV\big[&(\mathcal{A}/2)(\nabla\mathbf{n})^2-(K/2) n_z^2\nonumber\\
&+\mathcal{F}_D + \mathcal{F}_{mec}\big]
\end{align}
where $\mathcal{A},a$ and $K$ are the inhomogeneous, homogeneous
exchange and uniaxial anisotropy constants, respectively\cite{bogdanov1986phase,tveten2016intrinsic,affleck1989quantum,qaiumzadeh2018controlling}. The bulk DMI Energy density  may be written as $\mathcal{F}_D=\mathrm{D}\mathbf{n}\cdot(\nabla\times\mathbf{n})$ which leads to Bloch skyrmions in non-centrosymmetric materials such as MnSi and FeGe \cite{muhlbauer2009skyrmion,yu2011near,tonomura2012real}. as well as, the form of $\mathcal{F}_D=\mathrm{D}[(\mathbf{n}\cdot\hat{z})(\nabla\cdot\mathbf{n})-(\mathbf{n}\cdot\nabla)(\mathbf{n}\cdot\hat{z})]$ may create the  N\'{e}el skyrmion\cite{kezsmarki2015neel,fert2013skyrmions}. The DM interaction, being responsible for chiral spin structure, exists in Nio and CoO AFMs \cite{guo2022effect,akanda2020interfacial,kuswik2019determination}. The significance of the mentioned materials will be cleared later. The last term is the ME coupling as follows,
\begin{align}\label{eq:3 std Fmec}
 \mathcal{F}_{mec}=&\mathcal{F}^{\mathrm{sta}}_{mec}+\mathcal{F}^{\mathrm{std}}_{mec}\nonumber\\
 =&\sum_{i,j}\undertilde{\mathbf{B}}_{(1,2)}\mathbf{n}_i~{\bm{\undertilde\epsilon}^{\rm{sta}}_{~ij}}\mathbf{n}_j+\sum_{i,j}\undertilde{\mathbf{B}}'_{(1,2)}\mathbf{n}_i~{\boldsymbol{\undertilde\epsilon}^{\rm{std}}}\mathbf{n}_j
\end{align}
where the $\undertilde{\mathbf{B}}^{(')}_{(1,2)}=\mathrm{B}^{(')}_{1}\delta_{ij}+\mathrm{B}^{(')}_{2}(1-\delta_{ij}),\mathbf{n}_{i,j},{\bm{\undertilde\epsilon}^{\rm{sta}}_{ij}}, \boldsymbol{\undertilde\epsilon}^{\rm{std}} $, are the coupling tensor, magnetization in sites $(i,j)$, standard  and structure-dependent ME coupling, respectively. The first term is ubiquitous in both FM-AFM materials, however, there are some AFM constructions, such as $\mathrm{NiO}$ and $\mathrm{CoO}$ that have surplus ME term depending on intraplane spin structure \cite{simensen2019magnon} being introduced by the second term. The $i,j$ index is dropped from the second ME coupling since its elements are the combination of the standard $\epsilon_{ij}=\varepsilon^{\circ}_{ij}\sin(\mathrm{k}x-\omega t)$.
 In such a materials some neighboring atoms simultaneously belonging to both  sublattices alter the atmosphere. On the other hand, planes with internal ferromagnetic order in direction of a particular axis inside the net antiferromagnetic structure result in the additional ME strain tensor. Therefor, the ME coupling may be renormalized as\cite{simensen2018magnetostriction}:
\begin{align}\label{eq:4 renormalized mec}
&\mathcal{F}_{mec}=\begin{pmatrix}
\mathrm{n}_x\\
\mathrm{n}_y\\
\mathrm{n}_z
\end{pmatrix}^{\rm{T}}
\Biggl[\undertilde{\mathbf{B}}^{(1,2)}\begin{pmatrix}
\epsilon_{xx} & \epsilon_{xy} & \epsilon_{xz}\\
\epsilon_{yx} & \epsilon_{yy} & \epsilon_{yz}\\
\epsilon_{zx} & \epsilon_{zy} &\epsilon_{zz}
\end{pmatrix}^{\rm{sta}}+\nonumber\\
\rm{B'_1}&\begin{pmatrix}
(\epsilon_{xy}+\epsilon_{xz})&0&0\\
0&(\epsilon_{xy}+\epsilon_{yz})&0\\
0&0&(\epsilon_{xz}+\epsilon_{yz})
\end{pmatrix}^{\mathrm{std}}+\nonumber\\
\rm{B'_2}&\begin{pmatrix}
0~(\epsilon_{xx}+\epsilon_{xz}+\epsilon_{yy}+\epsilon_{yz})~(\epsilon_{xx}+\epsilon_{xy}+\epsilon_{zz}+\epsilon_{yz})\\
(\epsilon_{xx}+\epsilon_{xz}+\epsilon_{yy}+\epsilon_{yz})~0~(\epsilon_{yy}+\epsilon_{xy}+\epsilon_{zz}+\epsilon_{xz})\\
(\epsilon_{xx}+\epsilon_{xy}+\epsilon_{zz}+\epsilon_{yz})~(\epsilon_{yy}+\epsilon_{xy}+\epsilon_{zz}+\epsilon_{xz})~0\nonumber\\
\end{pmatrix}^{\mathrm{std}}\Biggr]\\
&\begin{pmatrix}
\mathrm{n}_x\\
\mathrm{n}_y\\
\mathrm{n}_z
\end{pmatrix}.
\end{align}
Here we explained an extra interaction in the free energy. Then by the AFM Lagrangian, $\mathfrak{L}[\mathbf{n}]=\int{\big[(1/2a)\mathbf{\dot{n}}^2}-(\mathcal{A}/2)(\nabla\mathbf{n})^2-\mathcal{F}_{mec}\big]~d^2\mathbf{r}$, and Rayleigh dissipation function, $\mathcal{R}(\dot{\mathbf{n}})=(\alpha_G\mathrm{M}_s/2\gamma)\int\mathbf{\dot{n}}^2d^2\mathbf{r}$ , we calculate the skyrmion equation of motion as, $8\pi\mathcal{M}_{ij}\bigg(\ddot{\xi}_j+\frac{a\alpha_G\mathrm{M}_0}{\gamma}\dot{\xi}_j \bigg)+ \mathbf{F_{\mathrm{tot}}(\boldsymbol{\xi}}(t))=0$. Based on these equations, Eqs. \ref{eq:5 Lagrangian}, \ref{eq:6 Rayleigh} and \ref{eq:8 coordinate EOM}, we describe main result of our findings briefly in the next subsection. \\
\subsection{main results}
We just explained the model on which we establish our theory and calculations. Equation .\ref{eq:12 EOM} explains the equationof motion of the soliton in the structure-dependent AFM. Its derivation details have been written in the appendix. It is a non-linear and non-homogeneous second order differential equation that could not be solved analytically. Therefore using same set of parameters as before, we simulate the equation to find center of mass motion for two different cases of $\beta$. Although the skyrmion with $\beta=0$ have ubiquitously been observed, there are hybrid skyrmions \cite{akhir2024stabilization} with $\beta=\pi/4$. For an effective one-dimensional system, we consider the SAW as $\epsilon_{xx}=\varepsilon_{xx}^{\circ}\sin(\mathrm{k}x-\omega t)$.
\begin{figure}[!t]
	\includegraphics[width=9.0cm,height=4cm,]{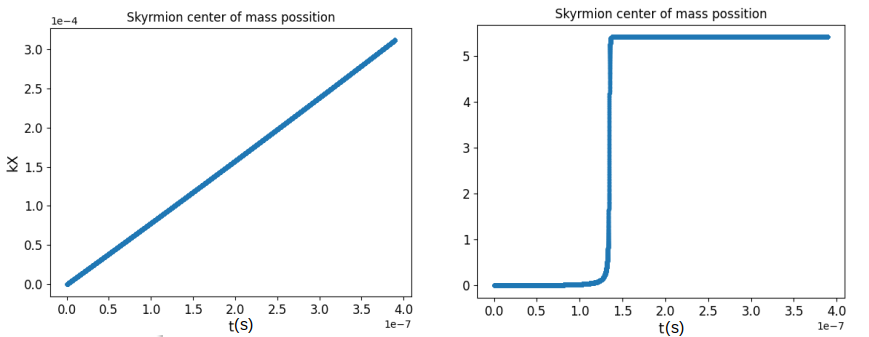}
	\caption{ Center of mass motion of skyrmion under the effect of SAW for (a) $\beta=0$ . For a time-scale of order 100 ns, there is a slow motion from initial position. Center of mass motion of skyrmion under effect of standing SAW for $\beta=\pi/4$. We can observe a strong motion toward antinodes of SA waves in the presence of $\rm{B'}_2$. }
	\label{sky_motion}
\end{figure}
We solve equations for a time-scale of order 100 ns which is a realistic scale in study of this type of system. 
As displayed in FIG. \ref{sky_motion} (a), in $\beta=0$, we observe a slow motion  and the system remains, effectively, in the same position in realistic time-scale. On the contrary, in  case $\beta=\pi/4$, there is a clear translation towards a stationary position of induced force by SAW, FIG. \ref{sky_motion} (b). Here we demonstrate the structure-dependent coefficient $\rm{B'}_2$ has a significant impact on skyrmion dynamics. This type of motion can be found in other closely related systems \cite {nepal2018magnetic}. Magnitude and time-scale of this motion depend on different parameters, e.g. effective damping, force strength and skyrmion radius, but general behavior of the system remains the same for wide range of parameters. To solve the Eq. \ref{eq:12 EOM} we recall the magnitudes used in typical materials from the following references, \cite{baltz2018antiferromagnetic,gomonay2014spintronics,jungwirth2018multiple,jungwirth2016antiferromagnetic,macdonald2011antiferromagnetic,marti2015prospect}. The required magnitudes are as follows, $a=10^{20} \mathrm{AT/Nm},~ \alpha_G =0.01,~ \mathrm{M}_0=10^6 \mathrm{A/m},~ \gamma =1.7\times 10^{11}\mathrm{rad/Ts},~ \mathrm{B_1}=\mathrm{B'_2}=10^6 \mathrm{J/m^3}, \varepsilon^{\circ}_{xx}=10^{-4}, \mathrm{k}=10^7 m^{-1}, \mathrm{w}=10^{-9}$. These parameters may display a typical motion of skyrmion in the presence of structure-dependent magneto-elastic coupling.\\
In the next section we will study the skyrmion dynamic in the presence of magnetoelastic coupling with its extra term.
\section{skyrmion dynamics in C{\smaller o}O-like structures}\label{IV}
The dynamics state motion of skyrmion may be extracted from Thiele approach\cite{thiele1973steady}. 
Hence, one would obtain the equation of motion for staggered magnetization using Lagrangian.
AFM Lagrangian, being the combination of 
the kinetic and free energy reads as following  in exchange approximation, \cite{tveten2014antiferromagnetic,shiino2016antiferromagnetic,hals2011phenomenology,tveten2013staggered,bender2017enhanced,velkov2016phenomenology}:
\begin{align}\label{eq:5 Lagrangian}
\mathfrak{L}[\mathbf{n}]=\int{\big[(1/2a)\mathbf{\dot{n}}^2}-(\mathcal{A}/2)(\nabla\mathbf{n})^2-\mathcal{F}_{mec}\big]~d^2\mathbf{r},
\end{align}
where $a^{-1}=\frac{\mathrm{M_s} }{\gamma^2 H_{exc}}$ is the inertia of AFM order parameter and the dot sign displays the time derivative as well as $ \mathrm{M_s}$ and $H_{exc}$ are the saturation magnetization and exchange field, respectively\cite{galkina2018dynamic,ivanov2005mesoscopic,tveten2016intrinsic} and the last two terms are total free energy, $\mathfrak{F}[\mathbf{n}]$, which is the functional of staggered magnetization.
The medium's viscous force is able to reduce the dynamics of magnetization then it must come in play. It is represented by Rayleigh dissipation function as,
\begin{align}\label{eq:6 Rayleigh}
\mathcal{R}(\dot{\mathbf{n}})=(\alpha_G\mathrm{M}_s/2\gamma)\int\mathbf{\dot{n}}^2d^2\mathbf{r},
\end{align}
where $\alpha_G$ is the Gilbert damping. The Euler-Lagrange minimization along with the $\mathrm{n}^2=1$ constrain lead to the following equation of motion,
\begin{align}\label{eq:7 staggered EOM}
\mathbf{n}\times\Big(~\ddot{\mathbf{n}}+a\frac{\delta\mathfrak{F}}{\delta\mathbf{n}} +(a\mathrm{M}_s\alpha_G/\gamma) ~\dot{\mathbf{n}}~\Big)=0.
\end{align}
In this level we are equipped to study the equation of motion of skyrmion in the presence of SAW. Here the additional term, relating to internal structure of $\rm{CoO} $, will revise the results. Consequently, we proceed to exploit the collective coordinate concept $\mathbf{n(r)}=\mathbf{n(r-\boldsymbol{\xi}}(t),t)$  to separate the time dependent softest modes and internal variable \cite{clarke2008dynamics}. Here the soft modes are $\boldsymbol{\xi}(t)=(\mathrm{X}(t)~,\mathrm{Y}(t))$ that the equation of motion is illustrated in terms of the soft modes. By multiplication of $\mathbf{n}\cdot\frac{\partial\mathbf{n}}{\partial\xi_i}\times$,  using $\mathbf{\dot{n}}=-\dot{\xi}_j\partial_{\xi_j}\mathbf{n}$ and $\mathbf{\ddot{n}}=-\ddot{\xi}_j\partial_{\xi_j}\mathbf{n}+\dot{\xi}_j\dot{\xi}_l\partial_{\xi_j}\partial_{\xi_l}\mathbf{n}$ and integration the skyrmion center of mass equation of motion, Eq. \ref{eq:7 staggered EOM}, changes to,
\begin{align}\label{eq:8 coordinate EOM}
&8\pi\mathcal{M}_{ij}\bigg(\ddot{\xi}_j+\frac{a\alpha_G\mathrm{M}_0}{\gamma}\dot{\xi}_j \bigg)+ \mathbf{F_{\mathrm{tot}}(\boldsymbol{\xi}}(t))+\mathcal{O}(\mathrm{N.L})=0,
\end{align}
so that,
\begin{align}\label{eq:9-10 mass tensor}
&\mathcal{M}_{ij}=\frac{1}{8\pi a}\int d^2\mathbf{r}~\frac{\partial\mathbf{n}}{\partial\xi_i}\cdot\frac{\partial\mathbf{n}}{\partial\xi_j}~,\\
&\mathbf{F_{\mathrm{tot}}(\boldsymbol{\xi}}(t))=-\int\label{eq:10 force} d^2\mathbf{r}\big(\frac{\delta\mathcal{F}^{\mathrm{sta}}_{mec}}{\delta\mathbf{n}}+\frac{\delta\mathcal{F}^{\mathrm{std}}_{mec}}{\delta\mathbf{n}}\big)\cdot \frac{\partial\mathbf{n}}{\partial\xi_i}
\end{align}
where $\mathcal{M}_{ij},~\mathbf{F_{\mathrm{tot}}(\boldsymbol{\xi}}(t))$ and $\mathcal{O}(\rm{N.L})$ are the skyrmion mass tensor, the total effective force and the ignored non-linear terms, respectively. Neglecting the non-linear terms, zero Magnus force in AFM \cite{barker2016static} and no edge repulsion insert simplifications to the issue. Another excessive postulate, being inevitable now, is the skyrmion motion similarity to that of a rigid body with no deformation. This fact prompts to the absence of D\"{o}ring mass \cite{doring1948tragheit} and conducts us to discus only upon the second term. This expression illuminating the mechanism of SAW induced motion of skyrmion contains the standard and structural contributions that will be figured out in the following. The unit vector of the staggered magnetization may be written as $\mathbf{n(r)}=\big(\sin\Theta(\rho)\cos\Psi(\varphi),\sin\Theta(\rho)\sin\Psi(\varphi),\cos\Theta(\rho)\big)$
so that, in cylindrical coordinate, $\Theta(\rho)$ and $\Psi=\varphi+\beta$ determine the texture profile and topological number \cite{nagaosa2013topological}
 with boundary values of $\Theta \big(\rho\longrightarrow(0,\infty)\big)=(\pi,0)$, with the unit vorticity and helicity, $\beta$. The mentioned prerequisites assist us to calculate the induced SAW force on skyrmion. 
Then we compute the effective force being usual in all kinds of AFMs, $\mathbf{F^{\rm{sta}}(\boldsymbol{\xi}}(t))$,  and the structure dependent contribution of the force, $\mathbf{F^{\rm{std}}(\boldsymbol{\xi}}(t))$.
Therefore the total ME force toward $\mathrm{Y}$ direction vanishes and for $\mathrm{X}$ direction, $\boldsymbol{\xi}(t)=\mathrm{X}(t)$,
 is the summation of $\mathbf{F^{\rm{sta}}}(\rm{X})$ and $\mathbf{F^{\rm{std}}}(\rm{X})$ that may be substituted in Eq. \ref{eq:8 coordinate EOM}.  Here the most significant terms affecting the skyrmion dynamics are the coefficients of $\epsilon_{xx}^{\circ}$. Therefor we ignore other terms and consider the time dependent polar component of magnetization introduced as a $360$\textdegree domain wall \cite{wang2018theory,braun1994fluctuations},
\begin{align}\label{eq:11 theta}
\Theta(\rho,t)=2\arctan\big[\frac{\sinh\frac{R}{\mathrm{w}}}{\sinh(\frac{\rho-\mathrm{X}(t)}{\mathrm{w}})}\big]
\end{align}
in which $R,w,\rho$ and $\mathrm{X}(t)$ are the radius, thickness of skyrmion wall, internal coordinate of the lattice atoms  and position of skyrmion  center, respectively. To solve the Eq.\ref{eq:8 coordinate EOM} one should compute the Eq.\ref{eq:10 force}. To do so, we call a mathematical approach utilized in \cite{wang2018theory} so that, $\Theta'\sin2\Theta\approx\frac{4}{\mathrm{w}}\delta(\mathrm{x}-g)$ and $\sin^2\Theta\approx 2\delta(\mathrm{x}-g)$ where $\mathrm{x}=\frac{\rho-\mathrm{X}(t)}{\mathrm{w}}$ and $g=\frac{R}{\mathrm{w}}$. Substituting them in Eq.\ref{eq:10 force} , then Eq.\ref{eq:8 coordinate EOM} , with no skyrmion deformation, could be rewritten as following,
\begin{align}\label{eq:12 EOM}
&\mathcal{M}\bigg(\ddot{\tilde{\mathrm{X}}}(t)+\frac{a\alpha_G\mathrm{M}_0}{\gamma}\dot{\tilde{\mathrm{X}}}(t) \bigg)=\nonumber\\
&\varepsilon^{\circ}_{xx}\frac{\mathrm{J}_2\big(\mathrm{k}\tilde{\mathrm{X}}(t)\big)}{\mathrm{k}}\big(1-\frac{\mathrm{w}}{\tilde{\mathrm{X}}(t)}\big)\Big( \mathrm{B_1}\cos2\beta+2\mathrm{B'_2}\sin 2\beta  \Big)\nonumber\\
-&\varepsilon^{\circ}_{xx}\mathrm{J_1}\big(\mathrm{k}\tilde{\mathrm{X}}(t)\big)\tilde{\mathrm{X}}(t)\Big( \mathrm{B_1}\cos^2\beta+\mathrm{B'_2}\sin 2\beta  \Big)
\end{align}\\
where $\tilde{\mathrm{X}}(t)=(R+\mathrm{X}(t))$ and $\mathrm{J_{1,2}\big(\mathrm{k}\tilde{\mathrm{X}}(t)\big)}$ are the first and second order Bessel functions.
\section{conclusion}\label{V}
We concluded that, isolated skyrmion may be nucleated by surface acoustic waves as an efficient low-energy method. This method is a viable substitution for other procedures being the energy-wasting as Joule heating. We deduced that the rate of skyrmion creation was proportional to amplitude of SAW. So that in the small amplitudes skyrmions could not be created any more, and topological number stayed zero. But by increasing the amplitude magnitude, skyrmions appeared and topological number tended to unity.\\
 Then, using Thiele approach, we derived a dynamical equation for skyrmion's motion in the specific type of AFMs. The $\rm{CoO}$ and $\rm{NiO}$ are the prominent examples of these type of AFMs. In these AFMs, because of the spin structure in $(111)$ parallel surfaces direction, the magnetoelastic interaction acquires excessive term that both depend on the structure of material and alter the soliton's dynamics. Based on the surplus structure-dependent term, we demonstrated it was crucial in $\rm{CoO}$-like materials. On the other hand, it could be able to bring new expressions for the dynamical equation of motion, specially in hybrid skyrmions. Numerical solution of EOM for $\beta=0$ and $\beta=\pi/4$ confirmed the significance of the structure-dependent term in a realistic time scales. Because in the former there were no effective motion but in the latter a motion took place from node to anti-node.
\section{Appendix}\label{VI}
\renewcommand{\theequation}{A.\arabic{equation}}
\setcounter{equation}{0}
Here we will write the details of Eq.\ref{eq:10 force} resulting in equation of motion Eq.\ref{eq:12 EOM}. Therefor we may write,
\begin{align}\label{eq:App force}
&\mathbf{F^{\rm{sta}}}\big(\mathrm{X}(t)\big)=-\int\mathrm d^2\mathbf{r}\frac{\delta\mathcal{F}^{sta}_{mec}}{\delta\mathbf{n}}\cdot\frac{\partial\mathbf{n}}{\partial \mathrm{X}},~ ~~\mathrm{X}=-x \nonumber\\
&=\int \mathrm{d}^2\mathbf{r}\big(\frac{\partial\mathcal{F}^{sta}_{mec}}{\partial\mathbf{n}_x}\frac{\partial\mathbf{n}_x}{\partial x}+\frac{\partial\mathcal{F}^{sta}_{mec}}{\partial\mathbf{n}_y}\frac{\partial\mathbf{n}_y}{\partial x}+\frac{\partial\mathcal{F}^{sta}_{mec}}{\partial\mathbf{n}_z}\frac{\partial\mathbf{n}_z}{\partial x}\big)\nonumber\\
&=\big(2\mathrm{B}_1\mathbf{n}_x \epsilon_{xx} + 2\mathrm{B}_2\mathbf{n}_y \epsilon_{xy} + 2\mathrm{B}_2\mathbf{n}_z \epsilon_{xz}\big)\nonumber\\
&\times\big(\Theta'(\rho) \cos\phi\cos\Psi(\phi)\cos\Theta(\rho) +\frac{1}{\rho}\sin\phi\sin\Theta(\rho)\sin\Psi(\phi) \big) \nonumber \\
&+\big(2\mathrm{B}_1\mathbf{n}_y \epsilon_{yy} + 2\mathrm{B}_2\mathbf{n}_x \epsilon_{xy} + 2\mathrm{B}_2\mathbf{n}_z \epsilon_{yz}\big)\nonumber\\
&\times\big(\Theta'(\rho) \cos\phi\cos\Theta(\rho)\sin\Psi(\phi) -\frac{1}{\rho}\sin\phi\sin\Theta(\rho)\cos\Psi(\phi) \big) \nonumber \\
&+\big(2\mathrm{B}_1\mathbf{n}_z \epsilon_{zz} + 2\mathrm{B}_2\mathbf{n}_y \epsilon_{yz} + 2\mathrm{B}_2\mathbf{n}_x \epsilon_{xz}\big)\big(-\Theta'(\rho) \cos\phi\sin\Theta(\rho)\big).
\end{align}
By entering the staggered magnetization, $\mathbf{n(r)}=\big(\sin\Theta(\rho)\cos\Psi(\varphi),\sin\Theta(\rho)\sin\Psi(\varphi),\cos\Theta(\rho)\big)$ the standard force may be written as,
\begin{align}
\mathbf{F^{\rm{sta}}}\big(\mathrm{X}(t)&\big)=\nonumber\\
&\mathrm{B}_1\int\Bigg[\epsilon_{xx}\big(\Theta'\sin2\Theta\cos^2\Psi\cos\varphi +\frac{\sin^2 \Theta}{\rho}\sin2\Psi \sin\varphi  \big)\nonumber\\
&+\epsilon_{yy}\big(\Theta'\sin2\Theta~\sin^2\Psi~\cos\varphi- \frac{\sin^2 \Theta}{\rho}\sin2\Psi \sin\varphi \big)\nonumber\\
&-\epsilon_{zz}\big( \Theta'\sin 2\Theta\cos\varphi \big)\Bigg]\mathrm{d}^2\mathbf{r}+\nonumber\\
&\mathrm{B}_2\int\Bigg[\epsilon_{xy} \big(\Theta'\sin 2\Theta~\sin 2\Psi~\cos\varphi-\frac{2\sin^2 \Theta}{\rho}\cos 2\Psi \sin\varphi \big)\nonumber\\
&+\epsilon_{xz}\big(\Theta'\cos 2\Theta\cos\Psi\cos\varphi  +\frac{\sin2 \Theta}{\rho}\sin\Psi \sin\varphi\big)\nonumber\\
&+\epsilon_{yz}\big(\Theta'\cos 2\Theta\sin\Psi\cos\varphi  +\frac{\sin2 \Theta}{\rho}\cos\Psi \sin\varphi\big) \Bigg]\mathrm{d}^2\mathbf{r}.
\end{align}

Inserting $\epsilon_{ij}=\varepsilon^{\circ}_{ij}\sin(\mathrm{k}x-\omega t)$ and integrating over the skyrmion surface lead to,

\begin{align}\label{APP_eq:12 force}
&\mathbf{F^{\rm{sta}}}\big(\mathrm{X}(t)\big)=\nonumber\\
&\mathrm{B}_1\int\rho~\mathrm{d}\rho~\Bigg[\varepsilon^{\circ}_{xx}\Big( \frac{2\pi\Theta'\sin2\Theta}{\mathrm{k}\rho}\big(\mathrm{k}\rho \rm{J_1}(\mathrm{k}\rho)\cos^2\beta- \rm{J_2}(\mathrm{k}\rho)\cos2\beta\big)\nonumber\\
&\qquad\qquad\qquad\quad+\frac{4\pi}{\mathrm{k}\rho^2}\sin^2\Theta~\mathrm{J}_2(\mathrm{k}\rho)\cos2\beta\Big)\nonumber\\
\nonumber\\
&\qquad\qquad+ \varepsilon^{\circ}_{yy}\Big( \frac{2\pi\Theta'\sin2\Theta}{\mathrm{k}\rho}\big(\mathrm{k}\rho \rm{J_1}(\mathrm{k}\rho)\sin^2\beta+ \rm{J_2}(\mathrm{k}\rho)\cos2\beta\big)\nonumber\\
 &\qquad\qquad\qquad\quad-\frac{4\pi}{\mathrm{k}\rho^2}\sin^2\Theta~\mathrm{J}_2(\mathrm{k}\rho)\cos2\beta\Big)\nonumber\\
&\qquad\qquad-\varepsilon^{\circ}_{zz}~2\pi\Theta'\sin2\Theta \rm{J_1}(\mathrm{k}\rho)
\Bigg]+\nonumber\\
\nonumber\\
&\quad\mathrm{B}_2\int\rho~\mathrm{d}\rho~\Bigg[\varepsilon^{\circ}_{xy}\Big( \frac{2\pi\Theta'\sin2\Theta}{\mathrm{k}\rho}\big(\mathrm{k}\rho \rm{J_1}(\mathrm{k}\rho)-2\rm{J_2}(\mathrm{k}\rho)\big)\nonumber\\
&\qquad\qquad\qquad\quad+\frac{8\pi\sin^2\Theta}{\mathrm{k}\rho^2}~\mathrm{J}_2(\mathrm{k}\rho)\Big) \sin2\beta\Bigg].
\end{align}

This is the customary force that is obvious in a bipartite AFMs, now we obtain the force comes out in structure-dependent configurations by the same way,

\begin{align}
&\mathbf{F^{\rm{std}}}\big(\mathrm{X}(t)\big)=\nonumber\\
&\mathrm{B}'_1\int\rho~\mathrm{d}\rho~\Bigg[\Big( \frac{2\pi\Theta'\sin2\Theta}{\mathrm{k}\rho}\big(\mathrm{k}\rho \rm{J_1}(\mathrm{k}\rho)\cos^2\beta- \rm{J_2}(\mathrm{k}\rho)\cos2\beta\big)\nonumber\\
&\qquad\qquad\qquad\quad+\frac{4\pi}{\mathrm{k}\rho^2}\sin^2\Theta~\mathrm{J}_2(\mathrm{k}\rho)\cos2\beta\Big)(\varepsilon^{\circ}_{xy}+\varepsilon^{\circ}_{xz})\nonumber\\
&\qquad\qquad+\Big( \frac{2\pi\Theta'\sin2\Theta}{\mathrm{k}\rho}\big(\mathrm{k}\rho \rm{J_1}(\mathrm{k}\rho)\sin^2\beta+ \rm{J_2}(\mathrm{k}\rho)\cos2\beta\big)\nonumber\\
&\qquad\qquad\qquad\quad-\frac{4\pi}{\mathrm{k}\rho^2}\sin^2\Theta~\mathrm{J}_2(\mathrm{k}\rho)\cos2\beta\Big)(\varepsilon^{\circ}_{xy}+\varepsilon^{\circ}_{yz})\nonumber\\
&\qquad\qquad-~2\pi\Theta'\sin2\Theta \mathrm{J_1}(\mathrm{k}\rho)(\varepsilon^{\circ}_{xz}+\varepsilon^{\circ}_{yz})
\Bigg]\nonumber\\
&\quad+\mathrm{B}'_2\int\rho~\mathrm{d}\rho~\Bigg[\Big( \frac{2\pi\Theta'\sin2\Theta}{\mathrm{k}\rho}\big(\mathrm{k}\rho \rm{J_1}(\mathrm{k}\rho)-2\rm{J_2}(\mathrm{k}\rho)\big)\nonumber\\
&\qquad\qquad+\frac{8\pi\sin^2\Theta}{\mathrm{k}\rho^2}~\mathrm{J}_2(\mathrm{k}\rho)\Big) \sin2\beta(\varepsilon^{\circ}_{xx}+\varepsilon^{\circ}_{xz}+\varepsilon^{\circ}_{yy}+\varepsilon^{\circ}_{yz})\Bigg]
\end{align}

Then we only keep the coefficients of $\varepsilon^{\circ}_{xx}$ and insert the delta function instead of $\Theta'\sin2\Theta, ~~\sin^2\Theta$ as mentioned above. Finally we obtain the total force by integration over $\rho$,
\begin{align}\label{eq:12 total Force}
\mathbf{F_{\mathrm{tot}}}(\mathrm{X}(t))&=\mathbf{F^{\rm{sta}}}\big(\mathrm{X}(t)\big)+\mathbf{F^{\rm{std}}}\big(\mathrm{X}(t)\big)\nonumber\\
8\pi\mathrm{B_1}\varepsilon^{\circ}_{xx}&\Big(\tilde{\mathrm{X}}(t)~\mathrm{J_1}\big(\mathrm{k}\tilde{\mathrm{X}}(t)\big)\cos^2\beta\nonumber\\
&-\frac{\mathrm{J_2}\big(\mathrm{k}\tilde{\mathrm{X}}(t)\big)\cos2\beta}{\mathrm{k}}(1-\frac{\mathrm{w}}{\tilde{\mathrm{X}}(t)})\Big)\nonumber\\
8\pi\mathrm{B'_2}\varepsilon^{\circ}_{xx}&\Big(\tilde{\mathrm{X}}(t)~\mathrm{J_1}\big(\mathrm{k}\tilde{\mathrm{X}}(t)\big)\nonumber\\
&-\frac{2\mathrm{J_2}\big(\mathrm{k}\tilde{\mathrm{X}}(t)\big)}{\mathrm{k}}(1-\frac{\mathrm{w}}{\tilde{\mathrm{X}}(t)})\Big)\sin2\beta
\end{align}
\bibliography{AFM_Sky_text_edited.bib}
\end{document}